# Growth control of oxygen stoichiometry in homoepitaxial SrTiO$_3$ films by pulsed laser epitaxy in high vacuum

**Ho Nyung Lee**[1,*], **Sung S. Ambrose Seo**[1,2], **Woo Seok Choi**[1,3], **and Christopher M. Rouleau**[4]

[1]Materials Science and Technology Division, Oak Ridge National Laboratory, Oak Ridge, TN 37831, USA
[2]Department of Physics and Astronomy, University of Kentucky, Lexington, Kentucky 40506, USA
[3]Department of Physics, Sungkyunkwan University, Suwon 440-746, Korea
[4]Center for Nanophase Materials and Sciences, Oak Ridge National Laboratory, Oak Ridge, TN 37831, USA
*hnlee@ornl.gov

**ABSTRACT**

In many transition metal oxides (TMOs), oxygen stoichiometry is one of the most critical parameters that plays a key role in determining the structural, physical, optical, and electrochemical properties of the material. However, controlling the growth to obtain high quality single crystal films having the right oxygen stoichiometry, especially in a high vacuum environment, has been viewed as a challenge. In this work, we show that through proper control of the plume kinetic energy, stoichiometric crystalline films can be synthesized without generating oxygen defects, even in high vacuum. We use a model homoepitaxial system of SrTiO$_3$ (STO) thin films on single crystal STO substrates. Physical property measurements indicate that oxygen vacancy generation in high vacuum is strongly influenced by the energetics of the laser plume, and it can be controlled by proper laser beam delivery. Therefore, our finding not only provides essential insight into oxygen stoichiometry control in high vacuum for understanding the fundamental properties of STO-based thin films and heterostructures, but expands the utility of pulsed laser epitaxy of other materials as well.

## Introduction

Strontium titanate, $SrTiO_3$ (STO), is a widely studied perovskite material with a high dielectric constant at room temperature, making it suitable for high-voltage capacitors and microelectronic devices. When doped by cation substitution or adding oxygen vacancies, many exciting physical properties or phenomena emerge in STO. Examples include the observations of superconducting states below 0.35 K (Ref. [1]), high mobility[2], and blue light emission[3]. In addition, strain induced by heteroepitaxial growth on lattice mismatched substrates can make this material ferroelectric at room temperature[4]. Interestingly, strain-free STO has also been found to be ferroelectric for small cation deficiencies in nominally stoichiometric films[5]. When combined with other perovskites, STO is often used to fabricate heterostructures and superlattices that yield artificial materials with entirely new or enhanced properties, such as enhanced ferroelectric polarization and the formation of a two-dimensional electron gas with multichannel conduction due to the multivalent nature of Ti ions[6-9]. Lastly, STO is one of most widely used substrate materials for epitaxial synthesis of other perovskite-type materials due to a favorable lattice match[7,10-12].

For epitaxial synthesis of STO, pulsed laser epitaxy (PLE) has been one of the most widely used physical vapor deposition techniques[13-15]. Here, a focused laser pulse of high-energy photons ablates a target material and generates an energetic plume, thereby transferring vaporized species to a substrate. One of the strengths of PLE lies in its effective conveyance of cation stoichiometry from the target to the substrates, effectively enabling near-stoichiometric epitaxial thin-film growth. Due to its versatility in growing various thin films from inorganic materials such as oxides to even organic materials, PLE has been adopted for a wide range of materials research. However, PLE also has drawbacks as it is a highly energetic process[16,17], capable of creating micron-sized particulates and bombarding the substrate with ionic species having high kinetic energy. Although the particulate problem can be minimized by mechanical or geometrical means[18-20], the ionic bombardment effect also deserves consideration as it has been shown to play a significant role in determining film and interfacial quality[21,22]. Here, the energetic components of the plume not only influence the oxygen stoichiometry of the growing film itself, but also eject oxygen ions from the substrates or create a chemically broad interface when growth occurs in high vacuum[16,23,24]. Stoichiometry problems in PLE-grown STO have been highly debated in recent years due to the difficulty in obtaining quantitative information on oxygen concentration[21,23]. Although growth at high oxygen partial pressures is a relatively simple solution to compensate for oxygen loss, some perovskite oxides, e.g., $LaTiO_3$ (Ref. [9,25,26]) and $LaMnO_3$ (Ref. [27]), simply require low pressure for epitaxial growth. Moreover, when layer-by-layer growth is critical for sample quality and digital synthesis of artificial materials, e.g. epitaxial heterostructures and superlattices, the use of high oxygen partial pressure not only limits the choice of materials, but also hinders obtaining atomically abrupt interfaces because these pressures tend to yield three-dimensional or island growth.[28] Consequently, the inability to grow in low oxygen partials pressures whilst maintaining stoichiometry and sharp interfaces represents a serious obstacle in advancing our understanding on the physics of oxide thin films and interfaces.

In this paper, we report a new approach that pushes the boundary of PLE optimal growth toward conditions that were once believed impossible – i.e., the growth of *highly* insulating, *stoichiometric* STO films in *high* vacuum. This was achieved by controlling the kinetic energy of the plume and prolonging the nucleus ripening time to better incorporate atomic oxygen provided by the target.

## Results and discussions

Tuning the energetics of the laser plume was done by employing a projection beamline and changing the size of a rectangular aperture that was imaged onto the ablation target. This style of beam delivery produces a near top hat beam intensity profile at the target, and changing the aperture size has the effect of changing the size of the laser ablation spot on the target (A) while simultaneously keeping the fluence (*J*) nearly constant. This in turn was found to control the amount of oxygen vacancies in our epitaxial oxides. Using STO as a model system, we show that highly insulating STO films with a bulk-like lattice constant and negligibly small concentration (if not zero) of oxygen vacancies can be grown successfully by PLE in very high vacuum ($P_{O2}$ ~ $10^{-6}$ Torr).

Homoepitaxial STO films (50 nm in thickness) were synthesized using typical growth parameters ($J$ = 1.0 J/cm$^2$, laser repetition rate, $f$ = 2~5 Hz, substrate temperature, $T_s$ = 700 °C) under high vacuum ($P_{O2}$= $10^{-6}$ Torr) without any post oxidation processes to compensate potential oxygen vacancies. A KrF excimer laser ($\lambda$ = 248 nm) was used to ablate a single crystal STO target, and the TiO$_2$-terminated STO substrates were prepared using chemical and thermal processes. Typical surface qualities of such substrates are presented elsewhere[29]. Four different apertures with different sizes were used to image the laser beam, yielding different laser-beam spot sizes on the target, namely 7.7 (A1), 2.0 (A2), 1.1 (A3), and 0.4 mm$^2$ (A4). These spot sizes required approximately 7, 27, 60, and 200 laser pulses, respectively, to grow one monolayer (~0.4 nm) of STO as determined by real-time reflection high-energy electron diffraction (RHEED) monitoring of the specular spot. Note that conventional pulsed laser deposition to grow materials quickly utilizes spot sizes similar to A1 or A2 (Ref. [30]), but in order to impose atomic level control over the synthesis of heterostructures and superlattices, spot sizes similar to (A4) have proven optimum[7,9,24,26]. Note that regardless of the laser spot size, two-dimensional layer-by-layer growth occurred as evidenced by *in-situ* monitoring of the RHEED pattern and oscillation of the specular spot intensity (data not shown).

As shown in Figure 1, gradually decreasing the laser spot size from A1 to A4 produced a clear evolution of the STO film color, indicating a gradual change in oxygen vacancy concentration (an insulating STO single crystal is also shown for comparison). Note that STO films with oxygen vacancies have a bluish color while stoichiometric STO is transparent[24,31]. Figure 1b shows optical absorption spectra $\alpha(h\nu)$ for our samples, and note that as the dimensions of the laser spot were decreased, a systematic reduction in optical absorption was noted in the mid-infrared and visible regions of the spectra. This was due to decreasing Drude (free-electron-carrier) absorption ($\lambda$ > 1000 nm) and reduction of deep-impurity bands (387 nm < $\lambda$ < 1000 nm), respectively. The sudden increase in $\alpha$ at 387 nm (3.2 eV in photon energy) originates from the $O_{2p}$-to-$Ti_{3d}$ charge transfer transition in STO films and substrates, indicating the bandgap of STO. The two absorption bands located at 729 and 517 nm (1.7 and 2.4 eV in photon energy) are typical for reduced STO bulk crystals. The absorption peak at 729 nm is known to be enhanced by reducing STO crystals further[32], while the one at 517 nm is independent of the free carrier concentration and might originate from the excitation of an electron trapped by an oxygen vacancy, i.e., in an $F_1$ center[33]. In either case, these absorption bands clearly indicate that the samples grown with conditions produced by apertures A1−A3 are oxygen deficient STO.

Conventional four-probe *dc* transport data also showed sheet resistance data (Figure 1c) that was consistent with the optical absorption spectra. Note that the A4 sample grown with the smallest laser spot size was highly insulating (R > 100 MΩ) despite being grown in highly reducing conditions. In contrast, other samples grown with larger laser-spot sizes exhibited a metallic ground state and corresponding optical features, indicating the crucial role of plume energetics on oxygen vacancy generation.

Temperature-dependent Hall measurement results (Figure 2) show that carrier concentrations of our samples vary nearly five orders of magnitude depending on the size of pulsed laser beam. For the metallic samples (A1−A3), sheet carrier densities ranged from ~$10^{17}$ cm$^{-2}$ for the largest spot to ~$10^{16}$ cm$^{-2}$ for a spot 7 times smaller (Figure 2a). As expected from the results shown in Figure 1, the larger the laser beam, the higher the carrier density, and in the extreme case of A4, no measurable conducting carriers were noted. While not the main point of this work, it is worth pointing out that the conducting STO films have extremely high mobilities that exceed 10,000 cm$^2$V$^{-1}$s$^{-1}$, which is highly comparable with doped STO samples prepared by molecular beam epitaxy[2]. Note that the mobilities of the metallic samples are similar to each other (Figure 2b), indicating that the system is carrier density controlled with conduction band curvature (i.e. effective mass) and degree of scattering having negligible influence on the conductivity of these samples. The negative and linear $R_{xy}(H)$ data typical for conventional metals are clearly observed from the conducting STO films (A1–A3) as shown in the inset of Figure 2b.

While intuitively it is hard to accept that growth of an oxide in high vacuum can result in a film with proper oxygen stoichiometry, the optical and *dc* transport data of our homoepitaxial STO films suggest it is possible with proper understanding of the strong dependence of the PLE plume kinetics on laser spot dimensions. Previously, we reported that the control of plume energetics is critical to controlling the interfacial quality of oxide thin films, and reduction of plume energy by inducing more scattering of ablated species with the background gas can significantly reduce the kinetic growth process[16]. In order to check the energetics of PLE plumes, we used an ion probe[16,34], which measures the coulomb flux of material travelling from the target to the substrate. The ion probe was positioned close to the substrate to monitor the plume kinetics during homoepitaxy of STO thin films in high vacuum with the four different apertures.

Figure 3 shows the influence of spot size on the plume. First, there is a reduction in the amount of ionized species ablated by a laser pulse with a smaller beam size as indicated by the decreased ion probe intensity (and area under the curve). Second, there is a significant reduction in the time-of-flight (TOF) and hence kinetic energy of the PLE plume by reducing the laser spot size with constant fluence. As shown in Figures 3a and b, as well as their summary in c, despite the keeping the other PLE parameters unchanged, the kinetic energy of the PLE plume was a strong function of the plume expansion dynamics produced by the laser spot size. For example, when we used the largest laser spot, A1, the TOF required for bulk (signal peak) of the laser ablated species to arrive on the substrate surface after a laser pulse was about 2.4 $\mu$s, or ~1.3 times shorter than the TOF produced with the smallest aperture, A4 (3.2 $\mu$s for $P_{O2}=10^{-6}$ Torr) (Figure 3b). Since kinetic energy goes as velocity squared, this TOF difference makes the kinetic energy of the latter PLE plume more than half the former. Such a drastic change is equivalent to what one can achieve by increasing the oxygen partial pressure by four orders of magnitude (see Figure 3c). Hence, the PLE plume created by a smaller laser beam size is significantly less energetic, inducing significantly less ionic bombardment effects on the STO film being grown as well as on the substrate itself. The main advantage of this approach is thus minimizing the damage to the film and achieving less (if not any) oxygen vacancies, thereby producing more stoichiometric thin films even in highly reducing conditions. Moreover, this growth control is highly advantageous for the growth of oxide heterostructures, in which the chemical sharpness is critical to induce strong interfacial coupling since the conventional high vacuum growth without such growth control is known to yield a chemically broad interface[16,23,24]. Note that while it is not the main point of this work, we expect that further studies to identify the threshold laser spot size below which the films can become insulating may provide insight into quantitatively understanding the correlation between the laser spot size and oxygen vacancy formation.

Furthermore, the use of very low growth rate promoted by a reduced laser spot has several advantages (as mentioned before, the typical number of laser pulses is 200 for growing one unit cell thick STO, which is about an order of magnitude slower process than conventional approaches), namely better growth control as observed by persistent RHEED oscillations, smoother surfaces, and better oxygen stoichiometry. In particular, we attribute the good oxygen stoichiometry under these conditions to an extended nucleus ripening time and the ability to incorporate more atomic oxygen from the target. A similar result was reported for growth of high $T_c$ cuprate thin films by using a lower laser repetition rate[30].

In order to evaluate the oxygen stoichiometry of our films, we employed x-ray diffraction (XRD) as the lattice constant change is a good indicator of the stoichiometry variation by shifts in cation and/or oxygen concentration[8,27,35,36]. Figure 4a shows XRD $\theta$-$2\theta$ scans of our samples around the 002 STO Bragg peak, and it has been reported that non-stoichiometric STO films have larger lattice constants than stoichiometric ones[8,35-38]. As the laser spot size decreases from A1 to A4, the deviation of the film peak with respect to the substrate peak shrinks, indicating better stoichiometry. Interestingly, the STO thin film grown with the smallest laser spot size (A4) showed no observable peak deviation, largely confirming that the film is nearly, if not perfectly, stoichiometric. XRD reciprocal space maps (RSMs) around the 114 STO reflection (Figures 4b and c) corroborated these results, and also confirmed the film was coherent with the substrate. The lack of Kiessig fringes in the XRD data, typical for stoichiometric thin films such as those grown by MBE[39,40], further confirmed the stoichiometric nature of our STO film grown with the smallest spot size. We also note that the use of a lower laser repetition rate to better oxidize thin films by extending the ripening time of adatoms on the substrate surface during the growth was not able to avoid oxygen vacancy formation when large laser spots were used. A large spot size also resulted in poor surface quality as evidenced by the deterioration of the RHEED intensity during the growth. A common observation from pulsed laser deposited films in high vacuum is that the lattice expansion seems to be quite larger than what is expected from the results of bulk materials. On the other hand, it was reported that reducing STO crystals by annealing in vacuum as high as 1000 $^o$C did not alter the lattice volume at all[41,42], indicating that oxygen vacancy itself does not change the lattice volume significantly. Thus, the large lattice expansion observed in most PLE grown samples indicates that the conventionally grown STO films in high vacuum contain not only oxygen vacancies, but complicated defects and complexes[41,43,44]. Nonetheless, it is worth stressing that the major contributor to the metallic transport and distinct dark bluish color from reduced STO films is oxygen vacancies based on our optical and transport data.

**Conclusion**

In summary, a growth control process to prepare oxygen vacancy free STO thin films in high vacuum was discovered. So far, crucial PLE growth parameters have been thought to simply include oxygen partial pressure, substrate temperature, target-to-substrate distance, laser repetition rate, and laser fluence. Here, we demonstrate the critical role of the laser beam spot size in synthesizing high quality single crystalline oxide thin films by PLE. Thus, some of controversies and poor reproducibility of pulse laser deposited samples among different research groups, despite the fact that the identical growth parameters were used, may be understood in terms the often overlooked laser spot size and its influence on the plume dynamics. In addition, it has been conventional wisdom in the oxide community that it is inevitable that oxygen deficient films will be produced in high vacuum, so few have explored this growth space for high quality films.  However, our experimental results provide new insight into obtaining the right oxygen stoichiometry in PLE grown films *even* under highly reducing conditions. Therefore, our discovery not only provides indispensable insight into the PLE process itself, but also expands the degree of freedom and utility of PLE for synthesizing oxide heterostructures and artificial materials in a wide range of oxygen pressures.

**Methods**
Epitaxial STO thin films (50 nm in thickness) were grown on $TiO_2$-terminated (001) STO substrates by pulsed laser epitaxy (KrF, $\lambda$ = 248 nm) at 700 $^o$C in vacuum. The laser fluence was fixed at 1.0 J/cm$^2$, and the laser repetition rate was 2 Hz for A1 and 5 Hz for A2-A4 samples. Reflection high-energy electron diffraction (RHEED) was used to monitor the surface structure and to control the film thickness with atomic-layer precision. Temperature dependent *dc* transport measurements were performed with a 14 T physical property measurement system (PPMS, Quantum Design Inc.). The optical absorption spectra were measured by a UV-Vis-NIR spectrophotometer (Cary 5000) at room temperature. *In-situ* ion probe measurements were used to characterize the energetics and dynamics of the laser plume. The sample structure and crystallinity were characterized by high-resolution four-circle XRD (X'Pert, Panalytical Inc.).

Acknowledgements
The authors would like to thank D. G. Schlom and G. Eres for helpful discussions. This work was supported by the U.S. Department of Energy, Office of Science, Basic Energy Sciences, Materials Science and Engineering Division. Optical measurements were conducted as a user project at the Center for Nanophase Materials Sciences, which is sponsored at Oak Ridge National Laboratory by the Scientific User Facilities Division, U.S. Department of Energy. WSC was in part supported by the Basic Science Research Program through the National Research Foundation of Korea funded by the Ministry of Science, ICT and Future Planning (NRF-2014R1A2A2A01006478).

**Figure captions**

**Figure 1.** Growth control of homoepitaxial STO films. (a) Photographs of STO films grown with four different apertures showing a wide variation of color depending on the aperture size. A STO substrate is also shown for comparison. The laser-beam spot sizes on the target are 7.7 (A1), 2.0 (A2), 1.1 (A3), and 0.4 mm$^2$ (A4) and require approximately 7, 27, 60, and 200 laser pulses, respectively, to grow one monolayer (~0.4 nm) of STO. (b) Optical absorption spectra for the four STO films. The STO film grown with A4 revealed identical optical transparency compared with the bare STO substrate, implying good stoichiometry. (c) Sheet resistance ($R_s$) data representing conducting (A1–A3) and insulating (A4) STO thin films. The resistance from the A4 sample was too high to measure as it exceeded the instrumental limit.

**Figure 2.** Temperature dependence of (a) the sheet carrier density $n_{sheet}$ and (b) the mobility $\mu$ of STO films grown with different spot sizes (A1–A4). The inset in (b) is the magnetic field dependent Hall resistance measured at 2 K, showing negative and linear $R_{xy}(H)$ typical for conventional metals.

**Figure 3.** Ion-probe signal measured as a function of $P(O_2)$ for (a) A1 = 7.7 mm$^2$ and (b) A4 = 0.4 mm$^2$. The vertical dotted line is an aid to the eye to show that ionized species arriving at the substrate are delayed when the smaller aperture (A4) is used. (b) Calculated kinetic energy of Sr arriving at the substrate surface as a function of $P(O_2)$. The use of smaller aperture resulted in a significant reduction of the kinetic energy and growth rate. Note that kinetic energy reduction minimizes the ionic bombardment effect while reduced growth rate extends the nucleus ripening time. These features are highly beneficial for the production of well-oxidized films during PLE.

**Figure 4.** XRD (a) $\theta-2\theta$ scans near the 002 reflection and (b) RSMs around the 114 peak for STO films grown with A1−A4. Note that the intense peaks are from STO substrates. The larger the laser spot size, the more the expansion of the out-of-plane lattice constant. When A4 was used, the substrate peak overlapped the film peak without any fringes, indicating that the stoichiometry of the film is identical to that of the substrate. XRD reciprocal space maps confirm that the films are strained coherently.

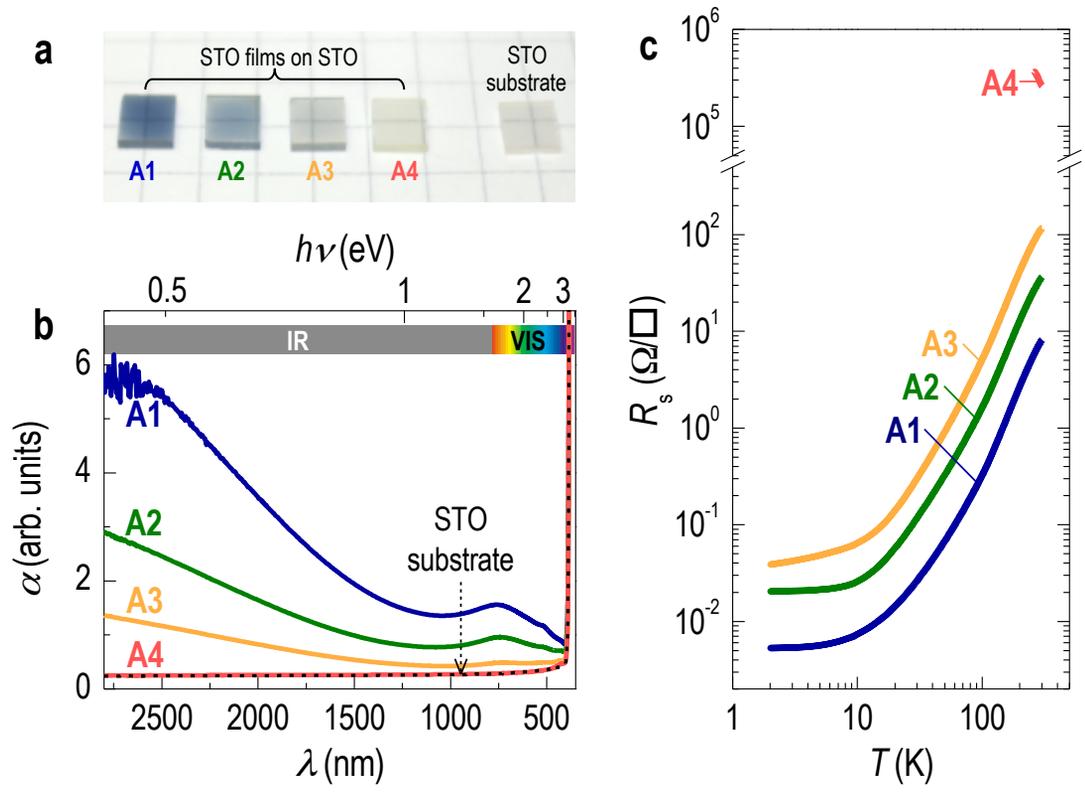

Fig 1

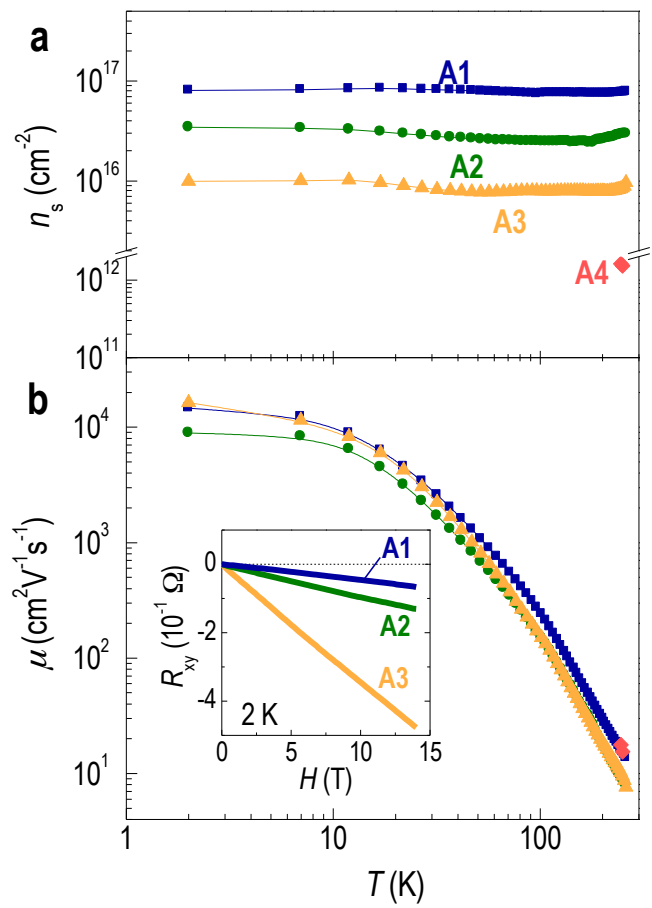

Fig 2

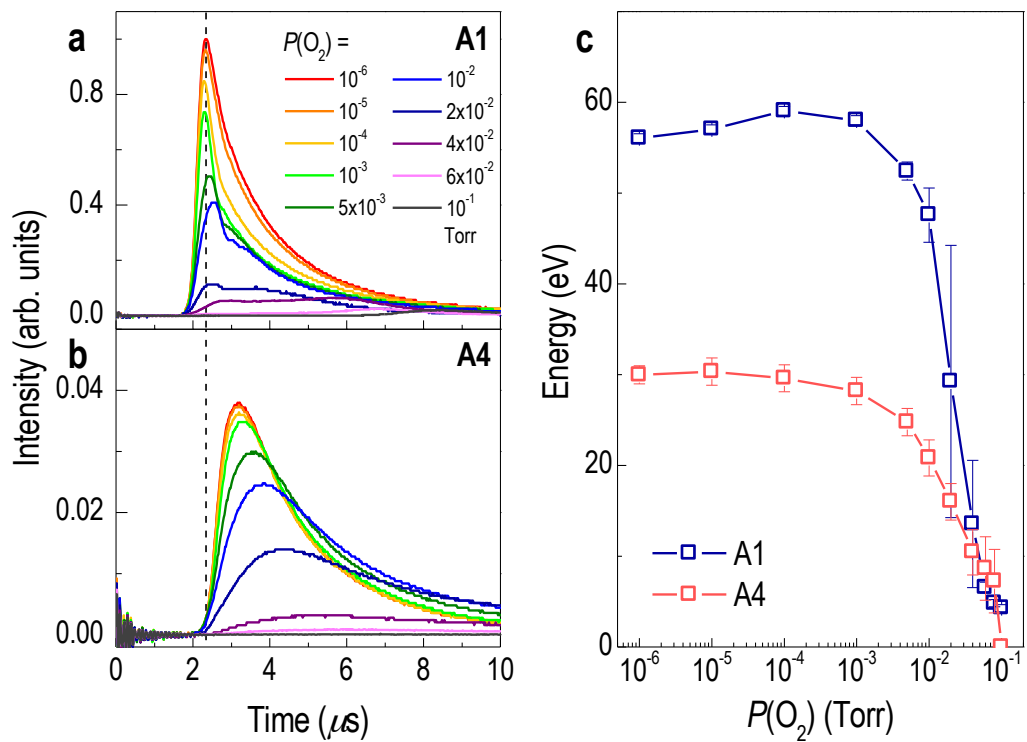

Fig 3

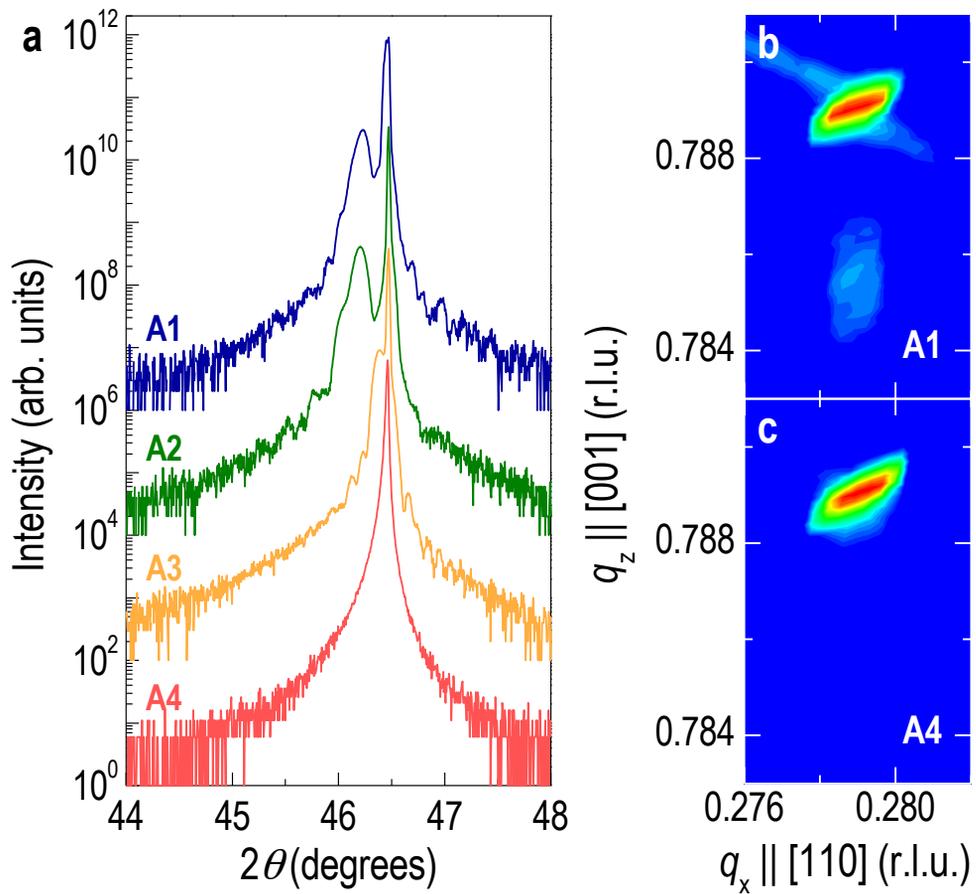

Fig 4